\documentclass[12pt]{article}
 \def\e{{\rm e}}
\newcommand{\be}{\begin{equation}}
\newcommand{\ee}{\end{equation}}
\newcommand{\bea}{\begin{eqnarray}}
\newcommand{\eea}{\end{eqnarray}}
\newcommand{\al}{\alpha}

\newcommand{\gm}{\gamma}
\newcommand{\Gm}{\Gamma}
\newcommand{\dl}{\delta}

\newcommand{\ep}{\epsilon}

\newcommand{\dd}{\mbox{d}}

\newcommand{\nn}{\nonumber}
\newcommand{\uk}{\underline{k}}
\newcommand{\Li}[2]{{\mbox{Li}}_{#1}\left(#2\right)}

\begin{document}
\parindent=1.5pc

\begin{titlepage}
\rightline{hep-ph/0007032}
\rightline{July 2000}
\bigskip
\begin{center}
{{\bf
Analytical Result for Dimensionally Regularized \\
Massless Master Double Box with One Leg off Shell
} \\
\vglue 5pt
\vglue 1.0cm
{ {\large V.A. Smirnov\footnote{E-mail: smirnov@theory.npi.msu.su.}
} }\\
\baselineskip=14pt
\vspace{2mm}
{\em Nuclear Physics Institute of Moscow State University}\\
{\em Moscow 119899, Russia}
\vglue 0.8cm
{Abstract}}
\end{center}
\vglue 0.3cm
{\rightskip=3pc
 \leftskip=3pc
\noindent The dimensionally regularized massless double box
{}Feynman diagram with powers of propagators equal to one, one leg
off the mass shell, i.e. with non-zero $q^2=p_1^2$, and three legs
on shell, $p_i^2=0,\;i=2,3,4$, is analytically calculated for general
values of $q^2$ and the Mandelstam variables $s$ and $t$. An explicit
result is expressed through  (generalized) polylogarithms, up to the
fourth order,  dependent on rational combinations of $q^2,s$
and~$t$, and a one-dimensional integral with a simple integrand
consisting of logarithms and dilogarithms.
\vglue 0.8cm}
\end{titlepage}

\section{Introduction}

Massless four point Feynman diagrams contribute to many important
physical amplitudes. They are much more complicated than two- and
three-point diagrams because depend on many parameters:
the Mandelstam variables $s$ and $t$ and the values of the
external momenta squared, $p_i^2,\; i=1,2,3,4$.
In the most general case, when all the legs are off the mass shell,
$p_i^2\neq 0$, there exists an explicit analytical result
\cite{UD} for the master (i.e. with powers of the
propagators equal to one) double box diagram (see Fig.~1)
\begin{figure}[hbt]
\centering
\begin{picture}(160,60)(0,2)
\put(20,10){\line(1,0){120}}
\put(20,50){\line(1,0){120}}
\put(40,10){\line(0,1){40}}
\put(80,10){\line(0,1){40}}
\put(120,10){\line(0,1){40}}
\put(40,10){\circle*{3}}
\put(40,50){\circle*{3}}
\put(120,10){\circle*{3}}
\put(120,50){\circle*{3}}
\put(80,10){\circle*{3}}
\put(80,50){\circle*{3}}
\put(6,9){$p_1$}
\put(6,49){$p_2$}
\put(146,9){$p_3$}
\put(146,49){$p_4$}
\put(32,27){\small $5$}
\put(72,27){\small $6$}
\put(112,27){\small $7$}
\put(59,0){\small $3$}
\put(99,0){\small $1$}
\put(59,53){\small $4$}
\put(99,53){\small $2$}
\end{picture}
\caption{}
\end{figure}
strictly in four dimensions.
Still no similar results are available for pure
off shell four point diagrams with ultraviolet, infrared and/or
collinear divergences.

In the opposite case, when all the end-points are on shell, i.e.
for $p_i^2=0,\; i=1,2,3,4$, the problem of the analytical evaluation
of such diagrams, in expansion in $\ep=(4-d)/2$
in the framework of dimensional regularization
\cite{dimreg} with the space-time dimension $d$ as a
regularization parameter,
was completely solved during last year in
\cite{K1,Smirnov:2000wz,Tausk}.
Among intermediate situations, when some legs are on shell
and the rest of them off shell, the case of one leg off shell,
$q^2=p_1^2 \neq 0$ and three legs on shell is very important
because of the relevance to the process $e^+e^-\to 3$jets
(see, e.g., \cite{3jets}). The purpose of this paper is to
analytically evaluate the master double box diagram of such
type, as a function of $q^2,s$ and~$t$, and thereby demonstrate
that the NNLO analytical calculations for this process are indeed
possible.

One of the ways to evaluate the four point diagrams with one leg
off shell is to expand them in the limit $q^2\to 0$ and compute
as many terms of the resulting expansion as possible. We explain
how to do this, following the strategy of regions \cite{BS,SR},
in the next section and present
the leading power term in this expansion which provides
a very non-trivial check of the subsequent analytical result.

To analytically evaluate the considered diagram
we straightforwardly apply the method of
ref.~\cite{K1}: we start from the
alpha-representation of the double box and, after expanding some
of the involved functions
in Mellin--Barnes (MB) integrals, arrive at a six-fold
MB integral representation with gamma functions in the integrand.
Then we use a standard procedure of taking residues and shifting
contours to resolve the structure of singularities in
the parameter of dimensional regularization, $\ep$.
This procedure leads to the appearance of multiple terms where
Laurent expansion
in $\ep$ becomes possible.
Resulting integrals in all the MB parameters
but the last two are evaluated explicitly in gamma functions
and their derivatives.
The last two-fold MB integral is evaluated by closing an initial
integration contour in the complex plane to the right,
with an explicit summation of the corresponding series.
A final  result is expressed through (generalized) polylogarithms  dependent
on rational combinations of $q^2,s$ and~$t$ and a
one-dimensional integral with a simple integrand consisting of
logarithms and dilogarithms.

\section{Expansion in the limit $q^2\to 0$}

The dimensionally regularized master massless double
box Feynman integral with one leg off shell, $q^2=p_1^2 \neq 0$, and
three legs on shell, $p_i^2=0,\; i=2,3,4$,
can be written as
\bea
{}F (s,t,q^2;\ep) &=&
\int\int \frac{\dd^dk \dd^dl}{(k^2+2 p_1 k+q^2) (k^2-2 p_2 k)
k^2 (k-l)^2}
\nn \\ && \times \frac{1}{
(l^2+2 p_1 l+q^2)(l^2-2 p_2 l) (l+p_1+p_3)^2 }
\, ,
\label{2box}
\eea
where $s=(p_1+p_2)^2, \;  t=(p_1+p_3)^2$, and
$k$ and $l$ are respectively loop momenta of the left and the right box.
Usual prescriptions, $k^2=k^2+i 0, \; s=s+i 0$, etc are implied.
To expand the given diagram in the limit $q^2\to 0$ one can
apply the so-called strategy of regions \cite{BS,SR} based on
the analysis  of various regions in the space of the loop integration
momenta, Taylor expanding the integrand in the parameters
that are considered small in the given region and
extending resulting integrations to
the whole integration domain in the loop momenta. When applying
this strategy all integrals without scale are by definition
put to zero.

Let us choose, for convenience, the external momenta as follows:
\[
p_1 = \tilde{p}_1 -\frac{q^2}{Q^2} \tilde{p}_2\,,\;\;\;
p_2=\tilde{p}_2\,,\;\;\;
\tilde{p}_{1,2} = (\mp Q/2,0,0,Q/2),
\]
where $s=-Q^2$.
The given limit $|q^2| \ll |s|, |t|$ is closely related to
the Sudakov limit so that
it is reasonable to consider each loop momentum to be one of
the following types:
\bea
\label{h}
\mbox{{\em hard} (h):} && k\sim Q\sim\sqrt{-t}\, ,
\nn \\
\label{1c}
\mbox{{\em 1-collinear} (1c):} && k_+ \sim q^2/Q,\,\,k_-\sim Q\, ,
\,\, \uk \sim \sqrt{-q^2}\,,
\nn \\
\label{2c}
\mbox{{\em 2-collinear} (2c):} && k_+\sim Q,\,\,k_-\sim q^2/Q\, ,
\,\,\uk \sim \sqrt{-q^2} \,.
\nn
\eea
Here $k_{\pm} =k_0\pm k_3, \; \uk=(k_1,k_2)$. We mean by $k\sim Q$, etc.
that any component of $k_{\mu}$ is of order $Q$.

It turns out that the (h-h), (1c-h) and (1c-1c) are the only non-zero
contributions to the leading power behaviour in the limit  $q^2\to 0$.
Any term originating from the (h-h) contribution is
given by the expansion of the integrand in Taylor series in $q^2$ and
expressed
through on-shell double boxes in shifted dimensions and can be
analytically evaluated by the algorithm presented in
\cite{Smirnov:2000wz}.
The (1c-1c)  contribution is obtained by expanding propagators
number~2, 4 and~7 in a special way. In particular, propagators
number 2 and 4 are expanded, respectively, in $l^2$ and $k^2$.
(See \cite{SR} for instructive 2-loop examples of expansions
in limits of the Sudakov type.)

The (1c-h) and (1c-1c) contributions are evaluated
with the help of a two-fold (respectively, one-fold) MB
representation. Still this program of the evaluation of a large number
of terms of the expansion looks very complicated because one needs,
for phenomenological reasons, the values of $q^2$ greater than
$s$ and $t$ so that a reliable summation of a resulting series,
using Pad\'{e} approximants, requires the knowledge of at least
first 20--30 terms. Such a great number of terms can be hardly
evaluated since a lot of irreducible structures appear.
This asymptotic expansion is however very useful for comparison
with the explicit result derived below.

The leading power terms of the asymptotic expansion calculated
in expansion in $\ep$, up to a finite part, are
\be
{}F (s,t,q^2;\ep) =
\frac{\left(i\pi^{d/2} \e^{-\gm_{\rm E}\ep} \right)^2 }
{(-s)^{2+2\ep}(- t)}
 \sum_{i=0}^4  \frac{ g_i (X,Y)}{\ep^i}
 + O(q^2 \ln^3 (q^2/s)) + O(\ep) \;,
\ee
where $X=q^2/s\,, \;Y=t/s$ and
\bea
g_4 (X,Y) &=&-1\;,
\nn \\   
g_3 (X,Y) &=&  -2(\ln X -\ln Y) \;,
\nn \\   
g_2 (X,Y) &=&
\frac{11\pi^2}{12} +3\ln X\,\ln Y -\frac{3}{2} \ln^2 Y \;,
\nn \\
g_1 (X,Y) &=&
2\ln Y\, \Li{2}{-Y}-2 \,\Li{3}{-Y}
+\frac{2}{3} \ln^3 X -\frac{3}{2} \ln^2 X \ln Y
-\frac{1}{2} \ln X\, \ln^2 Y
\nn \\ && \hspace*{-22mm}
-\frac{1}{6} \ln^3 Y  +\ln^2 Y\, \ln(1+Y)
+\pi^2 \left[ \frac{3}{2} \ln X-\frac{19}{6} \ln Y
+\ln(1+Y)\right]
+\frac{49\zeta(3)}{6} \;,
\nn \\
g_0 (X,Y) &=& 26\,\Li{4}{-Y}-2 S_{2,2}(-Y)
-2(\ln X+6\ln Y+\ln(1+Y))\,\Li{3}{-Y}
\nn \\ && \hspace*{-22mm}
+2\ln Y \,\Li{3}{\frac{Y}{1+Y}}
+( \ln^2 Y  +2\ln X \ln Y+4\pi^2) \,\Li{2}{-Y}
\nn \\ && \hspace*{-22mm}
-\frac{1}{2}\ln^4 X +\frac{1}{2} \ln^3 X \ln Y
+\frac{1}{4} \ln^2 X \ln^2 Y -\frac{1}{2} \ln X \ln^3 Y
+\frac{7}{8}\ln^4 Y
\nn \\ && \hspace*{-22mm}
+  \ln(1+Y) \left[
\ln X \ln^2 Y-\frac{5}{3} \ln^3 Y
+\frac{1}{2} \ln^2 Y\, \ln(1+Y)-\frac{1}{3} \ln Y\, \ln^2(1+Y)
\right]
\nn \\ && \hspace*{-22mm}
+\pi^2 \left[
-\frac{2}{3} \ln^2 X -\frac{7}{3} \ln X \ln Y
+\frac{25}{6}\ln^2 Y+\ln X \ln(1+Y) -2\ln Y \ln(1+Y)
\right.\nn \\ && \hspace*{-22mm}\left.
+\frac{1}{2}\ln^2(1+Y)
\right]
+\zeta(3)\left[
\frac{19}{3} \ln X-\frac{34}{3} \ln Y+2\ln(1+Y)\right]
+\frac{83\pi^4}{180} \;.
\label{LO}
\eea
Here $\Li{a}{z}$ is the polylogarithm \cite{Lewin} and
\be
\label{Sab}
  S_{a,b}(z) = \frac{(-1)^{a+b-1}}{(a-1)! b!}
    \int_0^1 \frac{\ln^{a-1}(t)\ln^b(1-zt)}{t} \dd t \;
\ee
the generalized polylogarithm \cite{GenPolyLog}.

\section{From alpha parameters through MB representation
to analytical result}

The alpha representation of the double box looks like:
\be
{}F (s,t,q^2;\ep) =
-\Gm(3+2\ep)\left(i\pi^{d/2} \right)^2
\int_0^\infty \dd\al_1 \ldots\int_0^\infty\dd\al_7
\dl\left( \sum \al_i-1\right) D^{1+3\ep}
A^{-3-2\ep} \; ,
\label{alpha}
\ee
where
\bea
D&=&(\al_1+\al_2+\al_7) (\al_3+\al_4+\al_5)
+\al_6 (\al_1+\al_2+\al_3+\al_4+\al_5+\al_7) \;, \\
A&=& [\al_1\al_2 (\al_3+\al_4+\al_5) + \al_3\al_4(\al_1+\al_2+\al_7)
+\al_6 (\al_1+\al_3)( \al_2+\al_4)](-s)
\nn \\&&
+ \al_5\al_6\al_7 (-t)
+\al_5 [ (\al_1+\al_3)\al_6 +\al_3(\al_1+\al_2+\al_7)] (-q^2)
 \; .
\eea
As it is well-known, one can choose
a sum of an arbitrary subset of $\al_i\,, i=1,\ldots,7$ in
the argument of the delta function in (\ref{alpha}), and we use
the same choice as in \cite{K1}.

Starting from (\ref{alpha}) we perform the same change of variables
as in \cite{K1}
and apply seven times the MB representation
\be
\frac{1}{(X+Y)^{\nu}} = \frac{1}{\Gm(\nu)}
\frac{1}{2\pi i}\int_{-i \infty}^{+i \infty} \dd w
\frac{Y^w}{X^{\nu+w}} \Gm(\nu+w) \Gm(-w)
\label{MB}
\ee
in order to separate terms in the functions involved to make
possible an explicit parametric integration. The two extra
MB integrations arise form the extra term with $q^2$.
After such integrations we are left with a 7-fold MB integral
of a ratio of gamma functions. Fortunately, one of the
integrations can be explicitly taken using the first Barnes
lemma and we arrive at the following nice 6-fold MB integral:
\bea
{}F (s,t,q^2;\ep) &=&
- \frac{\left(i\pi^{d/2} \right)^2}{\Gm(-1-3 \ep)(-s)^{3+2\ep}}
\frac{1}{(2\pi i)^6}
\int \dd v \dd w \dd w_2\dd w_3 \dd z\dd z_1
\left(\frac{q^2}{s} \right)^v
\left(\frac{t}{s} \right)^w
\nn \\ &&  \hspace*{-22mm}\times
\Gm(1+w) \Gm(1+v+w) \Gm(-v) \Gm(-w) \Gm(1-w_3+v)
\nn \\ &&  \hspace*{-22mm}\times
\Gm(w_2) \Gm(-1-2 \ep-w-w_2)\Gm(w_3-v) \Gm(-1-2 \ep-w-w_3)
\nn \\ && \hspace*{-22mm} \times
\frac{\Gm(1-w_2+z_1) \Gm(1-w_3+z_1) \Gm(\ep+w+w_2+w_3-z_1) \Gm(-z_1)}
{ \Gm(1+w+w_2+w_3) \Gm(-1-4 \ep-w-w_2-w_3) \Gm(1-w_3) }
\nn \\ && \hspace*{-22mm} \times
\Gm(1-\ep+z)
\Gm(2+2 \ep+w+w_2+z-z_1) \Gm(2+2 \ep+w+w_3+z-z_1)
\nn \\ && \hspace*{-22mm} \times
\frac{\Gm(-2-3 \ep-w-w_2-w_3+z_1-z) \Gm(z_1-z)}{ \Gm(3+2\ep+w+z)} \, .
\label{6MB}
\eea
It differs from its analog for $q^2=0$ by the additional integration in
$v$. This variable enters only four gamma functions in the
integrand.
The integral is evaluated in expansion in $\ep$, up to a finite
part, by resolving singularities in $\ep$ absolutely by the same
strategy as in the case $q^2=0$ \cite{K1}.
Note that the infrared and collinear poles are a little bit softer
than in the pure on-shell case, the integration variable $v$
playing the role of an infrared regulator.
The two key gamma functions that are responsible for the
generation of poles in $\ep$ are the same as in the previous
case:
\[ \Gm(\ep+w+w_2+w_3-z_1) \; \Gm(-2-3 \ep-w-w_2-w_3+z_1-z)\,. \]
The labeling of resulting terms is therefore the same:
the initial integral is decomposed as
$J=J_{00}+J_{01}+J_{10}+J_{11}$, etc. (Only arguments of some gamma
functions are shifted by $v$.)
The applied strategy makes it possible to perform all the
integrations apart from the last two, in $v$ and $w$.
We obtain four groups
of terms with 26 terms in each group: the terms without MB
integration, with MB integration in $v$ or $w$ and, finally,
with a two-fold integration in $v$ and $w$. The one-fold
integrals are explicitly evaluated by closing contour and summing up
series, using formulae from \cite{Oleg}.

The contribution of the resulting two-fold MB integral takes the form
\bea
\frac{2\left(i\pi^{d/2} \right)^2}{-s^{3}}
\frac{1}{(2\pi i)^2}\int \frac{\dd v \dd w}{1+w}
\left(\frac{q^2}{s} \right)^v
\left(\frac{t}{s} \right)^w
\Gm(1+v+w)\Gm(-v)\Gm(1+w)\Gm(-w)^2
&& \nn \\ && \hspace*{-135mm} \times
\left[
\Gm(1+v+w)\Gm(-v-w)
\left(
\frac{1}{\ep} -\gm_{\rm E} -2 \ln(-s)
-\frac{5}{1+w}-\frac{1}{1+v+w}
\right.\right. \nn \\ && \hspace*{-135mm} \left.
+\psi(1+v)-2\psi(-v-w)
-3\psi(-w)+2\psi(1+w)+\psi(1+v+w)
\right)
\nn \\ && \hspace*{-135mm} \left.
-\Gm(1+v)\Gm(-v) \Gm(1+w) \Gm(-w)
\right] \;.
\label{2MB}
\eea
The integration contours are straight lines along imaginary
axes with $-1<$Re$\,v$, Re$\,w$,Re$\,v+w<0$.
By closing contours it is possible to convert this integral into
a two-fold series where each term is identified as a derivative
of the Appell function $F_2$ in parameters, up to the third order.
The $1/\ep$ part is then explicitly summed up with a result in
terms of polylogarithms. (In fact, it is proportional to
the $\ep$ part of the master one-loop box.)

The so obtained result can be transformed into a one-dimensional
integral with a simple integrand. To present the final result
let us turn to the variables  $x=s/q^2$ and  $y=t/q^2$
keeping in mind typical phenomenological values of the involved
parameters relevant to the process $e^+e^-\to 3$jets:
\be
{}F (s,t,q^2;\ep) =
\frac{\left(i\pi^{d/2} \e^{-\gm_{\rm E}\ep} \right)^2 }{-s^2 t(-q^2)^{2\ep}}
 \sum_{i=0}^4  \frac{ f_i (x,y)}{\ep^i}
 + O(\ep)
 \;.
\ee
We obtain
\bea
f_4 (x,y) &=&-1\;,
\label{ResultEp4}
\\  
f_3 (x,y) &=&  2(\ln x +\ln y) \;,
\\   
f_2 (x,y) &=&
3 \,\Li{2}{x}+\,\Li{2}{y}-2(\ln x +\ln y)^2
\nn \\ &&
+3 \ln(1-x)\ln x + \ln(1-y)\ln y -\frac{5\pi^2}{12}
\;, \\
f_1 (x,y) &=&
2 \left[ \Li{3}{\frac{-x}{1-x-y}}+ \Li{3}{\frac{-y}{1-x-y}}
- \Li{3}{\frac{-x y}{1-x-y}}
\right.
\nn \\ && \hspace*{-20mm} \left.
- \ln x\, \Li{2}{\frac{y}{1-x}}- \ln y\, \Li{2}{\frac{x}{1-y}} \right]
+2 \ln(1-x-y)
\nn \\ && \hspace*{-20mm}
\times \left[
-\frac{1}{6} \left(\ln^2(1-x-y)+\pi^2 \right)
+  \ln(1-x)\ln x + \ln(1-y)\ln y - \ln x\ln y
\right]
\nn \\ && \hspace*{-20mm}
+3 \,\Li{3}{x}-8\,\Li{3}{y} +4 \,\Li{3}{\frac{-x}{1-x}}-2\,\Li{3}{\frac{-y}{1-y}}
\nn \\ && \hspace*{-20mm}
- (3\ln x+4\ln y) \Li{2}{x}+3\ln y \,\Li{2}{y}
+\frac{4}{3} \ln^3 x-\frac{2}{3}  \ln^3 (1-x)+ \ln^2 (1-x)\ln x
\nn \\ && \hspace*{-20mm}
-\frac{9}{2}  \ln (1-x)\ln^2 x +\frac{\pi^2}{6} (5\ln x-4\ln(1-x))
+\frac{4}{3} \ln^3 y+\frac{1}{3}  \ln^3 (1-y)
\nn \\ && \hspace*{-20mm}
-2 \ln^2 (1-y)\ln y
- \ln (1-y)\ln^2 y +\frac{\pi^2}{6} (5\ln y+2\ln(1-y))
\nn \\ && \hspace*{-20mm}
+4\ln x\, \ln y\, (\ln x-\ln(1-x)+\ln y)
+\frac{25\zeta(3)}{6} \;.
\eea
The $\ep^0$ part involves a one-dimensional integral:
\bea
f_0 (x,y) &=& \int_0^1 \dd z \left\{
z^{-1} \ln(1-z) (4 \ln^2(1-x -y z)- \ln^2(1-y-x z))
\right.
\nn \\ &&  \hspace*{-10mm}
- \frac{4 y}{1 - x -y z} \left[
\ln (1 - y z) \left(\ln(1 - z) \ln(1 - y z) - 2 \,\Li{2}{z}\right)
\right.
\nn \\ &&  \hspace*{-10mm} \left.
-2 (\ln(1 - z) - \ln z) 
\,\Li{2}{-(1 - x - y z)/x}
\right]
\nn \\ &&  \hspace*{-10mm}
- \frac{x}{1 - y -x z} \left[
\ln (1 - x z) \left(
3\ln^2(1 - z) -
6\ln(1 - z) \ln(1 - x z) + 2 \,\Li{2}{z}
\right)
\right.
\nn \\ &&  \hspace*{-10mm} \left.\left.
+2 (6\ln(1 - z) - \ln z)
\,\Li{2}{-(1 - y - x z)/y}
\right]
\right\}
\nn \\ && \hspace*{-22mm}
-5\,\Li{4}{x}+14\,\Li{4}{\frac{x}{1-y}}-2\,\Li{4}{\frac{-x}{1-x}}
-6\,\Li{4}{\frac{x y}{(1-x)(1-y)}}
\nn \\ && \hspace*{-22mm}
+8\,\Li{4}{\frac{-x}{1-x-y}}
+24\,\Li{4}{y}-2\,\Li{4}{1-y}+8\,\Li{4}{\frac{-y}{1-y}}-2\,\Li{4}{\frac{y}{1-x}}
\nn \\ && \hspace*{-22mm}
-8\,\Li{4}{1-x}-8\,\Li{4}{\frac{-y}{1-x-y}}
-20\,\Li{4}{\frac{1-x-y}{1-y}}+10\,\Li{4}{\frac{1-x-y}{1-x}}
\nn \\ && \hspace*{-22mm}
-3 S_{2,2}(x)-8 S_{2,2}(y)-6 S_{2,2}\left(\frac{x}{1-y} \right)
+( 2\ln y -2 \ln x -3\ln(1-x))\,\Li{3}{x}
\nn \\ && \hspace*{-22mm}
+2 (16\ln(1-y)-11\ln y -2\ln(1-x-y)+\ln x)\,\Li{3}{\frac{x}{1-y}}
\nn \\ && \hspace*{-22mm}
-(8\ln y+2\ln(1-x)+3\ln x) \,\Li{3}{\frac{-x}{1-x}}
\nn \\ && \hspace*{-22mm}
-2(4\ln(1-y)-4\ln y -\ln(1-x)+\ln x)
\Li{3}{\frac{x y}{(1-x)(1-y)}}
\nn \\ && \hspace*{-22mm}
+(14\ln(1-y)-18\ln y +4\ln(1-x-y))  \Li{3}{\frac{-x}{1-x-y}}
+2\ln y\, \Li{3}{\frac{-x y}{1-x-y}}
\nn \\ && \hspace*{-22mm}
+(7\ln y-8\ln (1-y)) \Li{3}{y}
+(8\ln (1-y)+\ln y +2\ln x) \Li{3}{\frac{-y}{1-y}}
\nn \\ && \hspace*{-22mm}
-4 (2\ln y +7\ln (1-x)+2 \ln(1-x-y) -8\ln x) \Li{3}{\frac{y}{1-x}}
\nn \\ && \hspace*{-22mm}
-2 (\ln y +5\ln (1-x)+8 \ln(1-x-y) -7\ln x) \Li{3}{\frac{-y}{1-x-y}}
\nn \\ && \hspace*{-22mm}
-\frac{1}{2} \left( \Li{2}{x}\right)^2
- \left(\Li{2}{\frac{x}{1-y}}\right)^2
-\frac{3}{2} \left(\Li{2}{y}\right)^2
+4 \left(\Li{2}{\frac{y}{1-x}}\right)^2
\nn \\ && \hspace*{-22mm}
+\left[
\ln^2 (1-x)-4\ln^2 y+2\ln y \,(4\ln(1-x)-\ln x)
-2\ln(1-y) \ln x-3\ln(1-x)\ln x
\right.\nn \\ && \hspace*{-22mm}\left.
 +\frac{7}{2}\ln^2 x+\frac{5\pi^2}{3}
\right] \Li{2}{x}
+\left[
12\ln^2 (1-y)+15\ln^2 y+2\ln(1-y) (\ln(1-x-y)+\ln x
\right.\nn \\ && \hspace*{-22mm}
-9\ln y)
+2\ln y \,(\ln(1-x-y)-4\ln(1-x) +4\ln x)
\nn \\ && \hspace*{-22mm}\left.
-2(\ln^2(1-x-y)+\ln^2 x) \right] \Li{2}{\frac{x}{1-y}}
+\left[-4\ln^2 (1-y)-11\ln^2 y
\right.\nn \\ && \hspace*{-22mm}\left.
+2\ln y \, (4\ln(1-x)-3\ln x)
+\ln(1-y) (5 \ln y-2\ln x) +\ln^2 x-\frac{\pi^2}{3}\right] \Li{2}{y}
\nn \\ && \hspace*{-22mm}
+\left[
8\ln^2 y-8\ln y\,\ln(1-x) -10\ln^2(1-x)
+8\ln^2(1-x-y)-8\ln(1-x-y)\ln x
\right.\nn \\ && \hspace*{-22mm}\left.
-8\ln(1-x) (
\ln(1-x-y)-2\ln x)+2\ln(1-y)\ln x
+\ln^2 x
-\frac{2\pi^2}{3} \right] \Li{2}{\frac{y}{1-x}}
\nn \\ && \hspace*{-22mm}
+\left[
\ln^2(1-x)-4\ln^2(1-y)-8\ln^2 y+2\ln y \, ( 4\ln(1-x)-3\ln x)
\right.\nn \\ && \hspace*{-22mm}\left.
 +2\ln(1-y) ( 4\ln y-\ln x)
-2\ln(1-x)\ln x +2\ln^2 x
\right] \Li{2}{\frac{x y}{(1-x)(1-y)}}
\nn \\ && \hspace*{-22mm}
+2\ln^4(1-x-y) +\frac{1}{3} \ln^3(1-x-y)\left[
2\ln y-3\ln(1-y)-9\ln x-11\ln(1-x)
\right]
\nn \\ && \hspace*{-22mm}
+\ln^2(1-x-y)\left[\pi^2-3\ln^2(1-y)+3\ln^2 y
+6\ln^2(1-x)-\ln(1-y) (\ln y-10\ln x)
\right.\nn \\ && \hspace*{-22mm}\left.
+4\ln(1-x)\ln x -2\ln^2 x -\ln y \,(5\ln(1-x)+\ln x)\right]
+\frac{1}{3} \ln(1-x-y)\left[7\ln^3(1-y)
\right.\nn \\ && \hspace*{-22mm}
-2\ln(1-y) (5\pi^2+12\ln^2 y) +7\pi^2\ln(1-x)
 +\pi^2 \ln x+6\ln^2 y \ln x-4\ln^3(1-x)
\nn \\ && \hspace*{-22mm}
+15\ln^2(1-y) (\ln y-2\ln x)
-21\ln^2(1-x)\ln x +3\ln(1-x)\ln^2 x
\nn \\ && \hspace*{-22mm}\left.
+\ln y \, (2\pi^2 +9\ln^2 x
+15\ln^2(1-x)-6\ln(1-x) \ln x)\right]
\nn \\ && \hspace*{-22mm}
-\frac{5}{6}\ln^4(1-x)-\frac{2}{3}\ln^4 x
+\frac{23}{6}\ln^3(1-x)\ln x-\frac{17}{4}\ln^2(1-x)\ln^2 x
+\frac{7}{2}\ln(1-x)\ln^3 x
\nn \\ && \hspace*{-22mm}
-\frac{\pi^2}{6}\left(
3\ln^2(1-x)-10\ln(1-x)\ln x+5\ln^2 x\right)
-\ln^4(1-y)-\frac{2}{3}\ln^4 y
\nn \\ && \hspace*{-22mm}
-\frac{19}{6}\ln^3(1-y)\ln y+5\ln^2(1-y)\ln^2 y
+\frac{2}{3}\ln(1-y)\ln^3 y
\nn \\ && \hspace*{-22mm}
+\frac{\pi^2}{6}\left(
9\ln^2(1-y)-\ln(1-y)\ln y-5\ln^2 y\right)
\nn \\ && \hspace*{-22mm}
+\frac{1}{3}\ln(1-x)\ln(1-y) (\ln^2(1-x)-4\ln^2(1-y))
-\frac{8}{3} (\ln^2 x+\ln^2 y)\ln x\,\ln y
\nn \\ && \hspace*{-22mm}
+3\ln^3(1-y)\ln x+
\ln^2(1-y) \left[\ln y (4\ln(1-x)+\ln x) -2\ln(1-x)\ln x\right]
\nn \\ && \hspace*{-22mm}
+\frac{1}{3}\ln y\, \left[-\ln^3(1-x)-9\ln^2(1-x)\ln x
-12\ln y\, \ln^2 x
\right.\nn \\ && \hspace*{-22mm}\left.
+6\ln(1-x)\ln x \, (2\ln y+3\ln x)\right]
-\ln(1-y) \left[8\ln^2 y \ln(1-x)
\right.\nn \\ && \hspace*{-22mm}\left.
+\ln(1-x) \ln x\, (\ln(1-x)-2\ln x)
+\ln y \, (-8\ln^2(1-x)+6\ln(1-x)\ln x +\ln^2 x)\right]
\nn \\ && \hspace*{-22mm}
+\frac{\pi^2}{3}\left[
\ln y \, (4\ln(1-x)-5\ln x)-\ln(1-y) \ln x \right]
\nn \\ && \hspace*{-22mm}
+\zeta(3) \left[
12 (\ln(1-x-y) -\ln(1-y)) +13\ln(1-x)
-\frac{25}{3}(\ln x+\ln y)
\right] +\frac{23\pi^4}{180} \,.
\label{Result}
\eea
One may hope that the one-dimensional integral that is left can
also be evaluated in terms of polylogarithms. To do this
it is necessary to complete the table of integrals derived in
\cite{Lewin}.

This result is in agreement with the leading power behaviour when
$q^2\to 0$ (\ref{LO}). When performing this comparison it is
reasonable to start with (\ref{2MB}), take minus residue at
$v=0$ (the first pole of $\Gm(-v)$), integrate in $w$ by closing
the contour to the right, and take into account the three other
contributions (without MB integration, and with integration
in $v$ or $w$) that were not presented above.
Eqs.~(\ref{ResultEp4}--\ref{Result}) also agree  with results based
on numerical integration in the space of alpha parameters \cite{BH}
(where the 1\% accuracy for the $1/\ep$ and $\ep^0$ parts is guaranteed).

\vspace{0.5 cm}

{\em Acknowledgments.}
I am grateful to Z.~Kunszt for involving me into this problem
and for kind hospitality during my visit to ETH (Z\"urich) in
April--May 2000 where an essential part of this
work was performed.
I am thankful to T.~Binoth and G.~Heinrich for comparison of the
presented result with their results based on numerical integration.
Thanks to A.I.~Davydychev and O.L.~Veretin for useful discussions.
This work was supported by the Volkswagen Foundation, contract
No.~I/73611, and by the Russian Foundation for Basic
Research, project 98--02--16981.

\end{document}